# About the possibility to observe diurnal variations of the count rate of dark photons by the multicathode counters.

A.V.Kopylov, I.V.Orekhov, V.V.Petukhov

Institute for Nuclear Research of RAS, 117312 prospect of 60 years of October 7A, Moscow, Russian Federation

The possibility to observe the diurnal variations in the count rate of dark photons using the multicathode counters is investigated. We show that placing detectors at three different mines at different latitudes: Pyhäsalmi in Finland, Baksan at Russia and INO in India and conducting the measurements with three different orientations of detectors one can obtain quite a complete series of data to determine the direction of the dark photon field. We present the results of the measurements fulfilled by the presently developed counters.

## INTRODUCTION

The nature of dark matter it is still unsolved mystery of modern physics. Following the astrophysical observation its mass by more than five times exceeds the mass of a luminous matter but still there haven't been obtained the reliable experimental data confirmed by any independent observation. Due to its high mass dark matter influences the motion of the stars but still it hasn't been detected by any physical apparatus despite enormous efforts of many researches to accomplish this task. At the present time the major work is focused on the search of weakly interacting massive particles (WIMPS). In elastic scatterings they produce an ionization track in a working medium of detector. In the most developed experiments they use the noble gases in a liquid phase [1 - 3]. The progress achieved in these experiments is so great that their upper limits are approaching the expected effect from solar neutrinos (neutrino floor). This sets a natural limit for further study by this method. It looks quite natural in these circumstances to extend the geography of the search. Among the multitude of the suggested theoretical models one can find also the one with the hidden-photons that has been proposed by L. B. Okun in 1982 [4]. The oscillations of hidden-photons into the real ones due to kinetic mixing [5] make possible the registration of this effect. The search of hidden-photons has been conducted in some experiments [6] and also in the ones using accelerators [7]. The new method of the search has been proposed in [8] using a dish antenna. This method has an upper limitation due to absorption of higher energy ultraviolet radiation by the material of antenna.

As an extension of this method to higher energy range we have proposed to use the effect of photoemission of electrons. In this case we should use a gaseous proportional counter with high ($>10^5$) gas amplification [9] to detect single electrons emitted from a cathode of the counter. In this detector we use free electrons of the degenerate electron gas of a metal as a target and the effect is proportional to the surface of the cathode. Because in this technique the electrons emitted from the surface of a metal are detected, it is sensitive for the energy of hidden photons greater than the work function for the metal of the cathode. Thus, this method extends the method of a dish antenna to the higher energy range typically greater than about 4 eV. In this range there are strong limitations obtained on detectors where the sensitive region is a volume of the detecting material [10, 11]. But, contrary to our case, the valence electrons have been used as a target in these experiments. Since we don't have a strict theory of hidden-photons by nowadays the physics of these two processes may be quite different. As it was shown in [8] the average power absorbed by antenna (in our case by a cathode of the counter) is described by

$$P = 2\alpha^2 \chi^2 \rho_{CDM} A_{cath} \qquad (1)$$

here $\alpha^2 = <\cos^2\theta>$, and $\theta$ – the angle between the field of hidden-photons and a vector normal to the surface of the cathode (the brackets denote the value averaged by a multitude of hidden-photons and the surface of a cathode ), $\rho_{CDM}$ - the local density of dark-photons, $A_{cath}$ – the surface of a cathode, $\chi$ – a kinetic mixing parameter that determines the "admixture" of photons and that is defined by the Lagrangian

$$L = -\frac{1}{4}F_{\mu\nu}F^{\mu\nu} - \frac{1}{4}F'_{\mu\nu}F'^{\mu\nu} - \frac{\chi}{2}F_{\mu\nu}F'^{\mu\nu} + \frac{m_{\gamma'}^2}{2}A_\mu A'^\mu \qquad (2)$$

where: $A_\mu$ and $A'_\mu$ are the photon and hidden-photon fields, $F_{\mu\nu}$ and $F'_{\mu\nu}$ - are field strengths respectively, and $m'_\gamma$ - is the hidden-photon mass.

As one can see from (1), the effect in our detector is proportional to the surface of the cathode. Cylindrical shape of the cathode enables to distinguish directions if the field of hidden-photons is not isotropic. For example, if the direction of the field is along the axes of the <, then $<\cos^2\theta> = 0$ and the effect is zero. The maximum of the effect will be when the direction of the field is perpendicular to the axes and $<\cos^2\theta> = ½$. The surface of the cathode should be quite smooth – mirror-like, because the surface irregularities will violate this equality. The counter with a mirror-like cathode can "see" direction, i.e. can work like a telescope. This can be used to get the evidence that hidden-photons have been really detected as it will be shown later. This paper has a following structure: first we describe our detector and discuss what should be done to fulfill this experiment, then we consider what data can be obtained by conducting measurements with three counters with orthogonal orientation placed in three different mines at different geographical latitudes: Pyhäsalmi in Finnland, Baksan in Russia and INO in India. In the final part of the paper we formulate conclusions.

## THE CONSTRUCTION AND WORK OF A MULTI-CATHODE COUNTER

At present time in INR RAS in Troitsk, Moscow the experiment PHELEX (PHoton-ELectron EXperiment) on the search of dark photons is conducted. The detailed description of the detector and the method of data treatment are presented in [9]. Here we give only short description of it that is necessary for further understanding. The method is based on the detection of single electrons emitted from the surface of a cathode as a result of the conversion of hidden-photon. In this case the power absorbed by a cathode can be found from the rate of counting of single electrons emitted from the cathode:

$$P = m_{\gamma'} R_{MCC} / \eta \qquad (3)$$

here: $m_{\gamma'}$ – mass of a hidden-photon, $R_{MCC}$ – the count rate of single electrons, $\eta$ – quantum efficiency to emit single electron that is taken here to be equal to the one for real photon with energy $E_\gamma = m_{\gamma'}$. Quantum efficiencies have been taken from [10]. We also assumed that all the energy density of dark matter is contained in hidden-photons, hence, the powers determined by (1) and (3) should be equal. Combining (1) and (3) we obtain expression for sensitivity of this method:

$$\chi = 2.9 \cdot 10^{-12} \left( \frac{R_{MCC}}{\eta \cdot 1Hz} \right)^{\frac{1}{2}} \left( \frac{m_{\gamma'}}{1eV} \right)^{\frac{1}{2}} \left( \frac{0.3 GeV/cm^3}{\rho_{CDM}} \right)^{\frac{1}{2}} \left( \frac{1m^2}{A_{cath}} \right)^{\frac{1}{2}} \left( \frac{\sqrt{2/3}}{\alpha} \right) \qquad (4)$$

As one can see from this expression the sensitivity of this method depends on the dark rate of counting which can be quite substantial due to edge effect, determined by the events with the tracks cut at the ends of the counter, due to emission of electrons from chemical impurities inside a counter and other sources. To subtract this background count rate from the total one we developed a counter of special construction. At a distance of about 5 mm from metallic cylindrical cathode 192 mm in diameter and 495 mm long we placed a second, internal one made of tighten nichrome wires 50 μm with a pitch 4.5 mm. The potential applied to this cathode was higher (in a first configuration) and lower (in the second configuration) than the one applied to outer cathode. In first configuration electrons emitted from a cathode could freely drift to a central anode wire where they were registered by preamplifier. In second configuration electrons were recoiled back by lower potential of second cathode and could not be registered. The measurements were conducted alternatively in these two configurations. In the first one the sum of the effect from electrons emitted from a solid cathode plus the background from other sources has been measured. In the second one only background has been measured in a very similar geometry. The effect has been found as a difference of the rates obtained in first and second configurations. At Fig.1 one can see the counter during assembly, at Fig.2 – a simplified scheme of the apparatus with a preamplifier and an 8-bit NI-5152 digitizer.

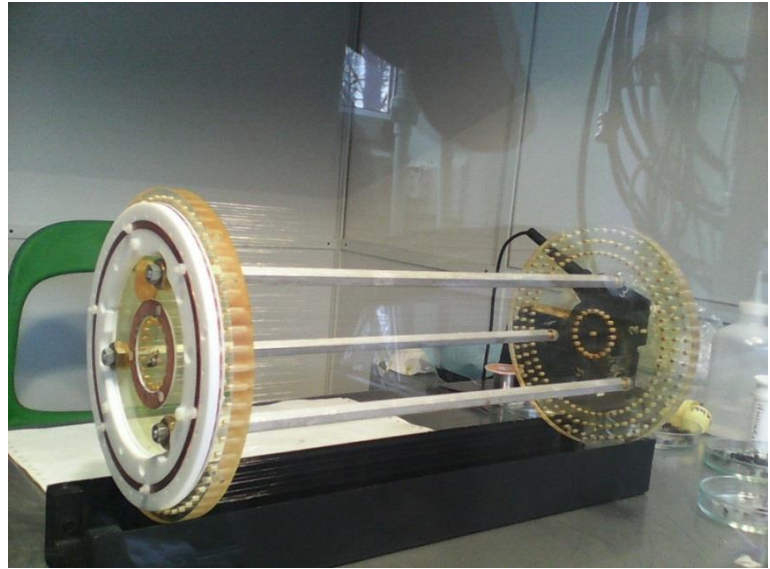

Fig. 1. Counter during assembly.

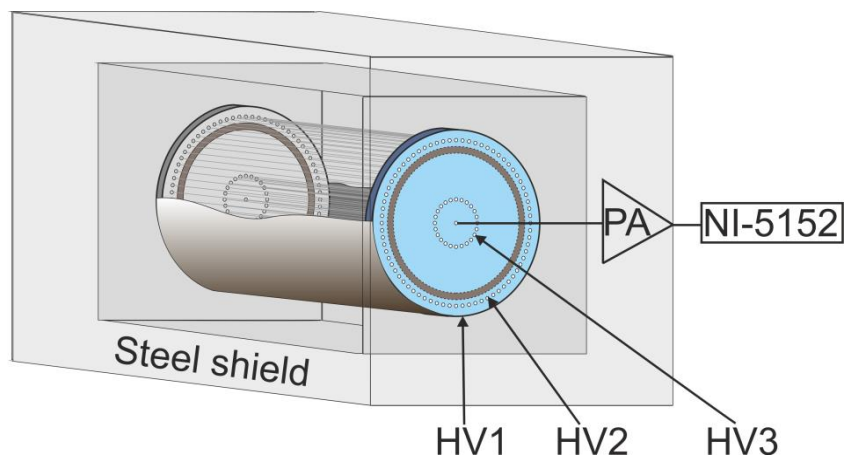

Fig. 2. A schematic of a multi-cathode counter. HV1, HV2, HV3 – high voltage applied to cathodes 1, 2 and 3. PA – charge sensitive preamplifier.

Around a central (anode) wire of 20 μm in diameter a third cathode was arranged made of tighten nichrome wires of 50 μm in diameter with a pitch 5 mm. The diameter of this cathode was 40 mm. High voltages HV1, HV2, HV3 were applied to cathodes to provide high ($>10^5$) gas amplification. Electrons emitted from first (solid) cathode drift towards anode wire where they produce an avalanche and are registered by charge sensitive preamplifier. The amplitude of the pulse is proportional to the charge collected on central wire. The shape of the pulse was recorded by an 8-bit ADC NI-5152 with a frequency 10 MHz in the interval ± 50 mV and a sampling step around 400 μV. Only the intervals were analyzed when the baseline did not deviate from zero beyond ±2 mV, only a live time was taken into account. The measurements have been conducted by series, 12 hours each: one at day time and one at night time. The results of each series were processed offline to select the true events by amplitude and pulse shape discrimination [9]. For each series we obtain one point presented at Fig. 3. The calibration was performed by UV-radiation of mercury lamp through a quartz window in the wall of the counter, the spectra of single electron events being recorded in both configurations. To shield from the surrounding gamma-radiation the counter was placed inside a steel cabinet with the walls of about 300 mm thick. The total assembly was placed at a ground floor of a building at Troitsk, Moscow.

Figure 3 shows the results of measurements obtained by using a counter with an aluminum cathode filled by gas mixture Ne+$CH_4$ (10%) at 0.1 MPa, with focusing rings at both ends of the counter (to reduce the background from tracks of muons cut at the ends of the counter).

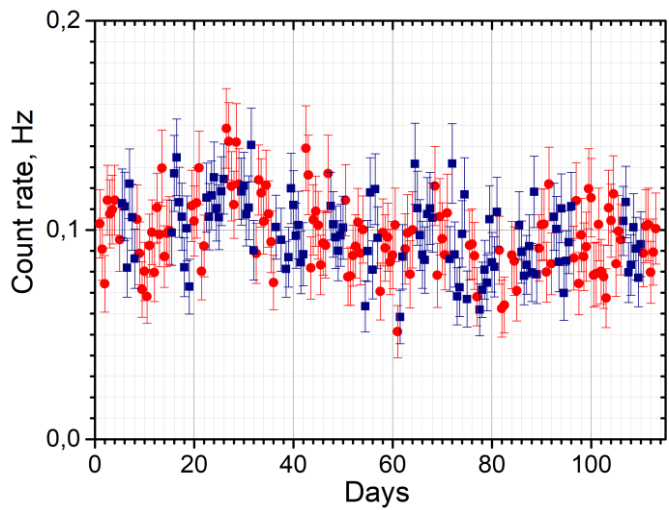

Fig. 3. The count rates measured in configurations 1 (circles) and 2 (squares). The temperature has varied in the region from 5ºC to 15ºC.

Subtracting the count rate in configurations 1 and 2 we obtain the count rate $r_{MCC} = R_{MCC}/A_{cath} = (-0.33 \pm 0.7) \cdot 10^{-6}$ Hz/см$^2$ that is the smallest of the rates of emission of single electrons measured by to-day. This is less than dark current measured by PMTs, EMs and other detectors what demonstrates the advantage of multi-cathode counter for this kind of measurements. Substituting $R_{MCC}$ in (4), we obtain an upper limit for parameter $\chi$. At Fig.4 one can see this upper limit at 95% CL obtained from measurements presented at Fig. 3 and also the one obtained earlier [14] with the same counter but with a mixture Ar + CH$_4$ (10%).

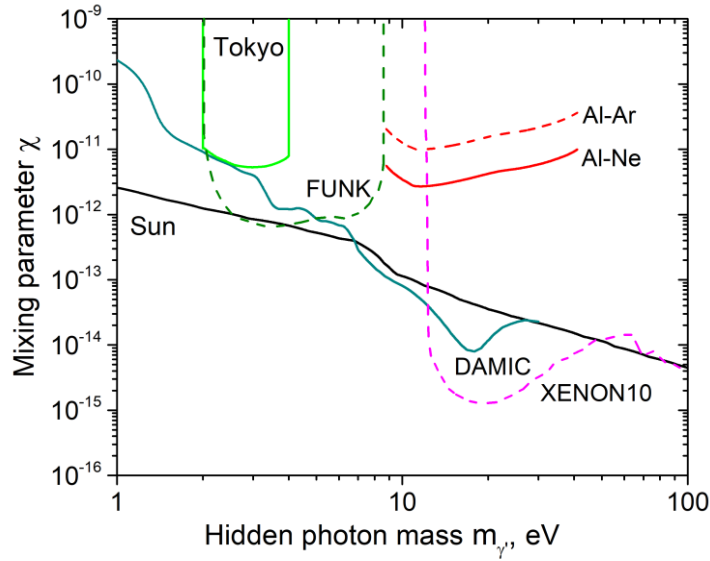

Fig. 4. The upper limits obtained in a series of measurements Al-Ar [14] with a mixture Ar + CH$_4$ (10%) and Al-Ne with a mixture Ne + CH$_4$ (10%). Here the upper limit Tokyo is taken from [13], FUNK – from [15], XENON10 from [11] and DAMIC – from [12].

As one can see from Fig.4, the upper limit obtained from our measurements is still in the region excluded by data from solar evolution. But here we should take into account that solar data are obtained for solar plasma with extreme densities and temperatures that are still not attainable in the laboratory. Thus, the physics of the processes in these two cases may differ in details that could be quite substantial. For this mass range there are also strict limitations obtained in experiments where as a target not a surface but a certain volume of a working medium is used. But in these detectors the target is the valence electrons while in our case the target is free electrons of a degenerate electron gas of a metal. To reduce further dark count rate it is necessary to subtract the background produced by muons which can't be neglected at this level of measurements or to place detector underground where the flux of muons is small. We are planning also to use materials with lower concentration of radioactive impurities and also to cool the detector to reduce thermionic dark count rate from impurities inside a detector. It is also possible to improve the result by increasing a live time of counting. Then it will be possible to increase the sensitivity of the method and to start the experiment described below.

**DIURNAL VARIATIONS FROM DARK PHOTONS IN THE MINES BNO IN RUSSIA, PYHÄSALMI IN FINLAND AND INO IN INDIA**

In experiments with a dish antenna [13], [15], and also in our measurements Al-Ar [14] and Al-Ne the direction of the field of dark photons has been assumed to be isotropic and in this case $\alpha^2 = <\cos^2\theta> = 2/3$. If the field of dark photons has the same direction everywhere in space it becomes possible to observe diurnal variations of the count rate of our detector due to variation of the value $\alpha^2 = <\cos^2\theta>$ throughout the day by the rotation of the Earth. For the observation of this effect it is expedient to have three counters of the same dimensions oriented orthogonal to each other. We can place one counter in vertical position; second one – along the parallel of the Earth and third one – along the meridian. Because the angle $\eta$ between vector of the field of dark photon and the axes of the counter will depend also from the geographical latitude of the place where detector is situated it looks attractive to choose three mines at different latitudes for the measurements: at high, medium and low latitudes. For the first one it is quite appropriate to use Pyhäsalmi mine in Finland at latitude 64º, for the second – Baksan mine (BNO) in Russian Federation at latitude 43º and for the third – INO mine in India at latitude 10º. At Fig/5 (a,b,c) the averaged within one hour values of $\alpha^2$ for each of these mines are given for three different orientations of the counter. One can see that diurnal variations of $\alpha^2$ are very characteristic for different orientations of the counters at different latitudes and for different directions of the field of dark photons relative to the Earth axes that will enable to find the angle $\eta$. The discovery of this difference in this experiment can be used as the evidence to prove the very fact of existence of dark photons with a field having a certain direction.

Here we should also take into account the difference in geographical longitude of these mines. For each mine we have taken the phase so that 0 hours corresponds to the moment when vector of the field of dark photons lies in the plane of the corresponding meridian. For each angle $\eta$ found from the temporal run of the count rates obtained for all nine detectors the absolute direction of the field of dark photons can be found from absolute phase of the temporal run.

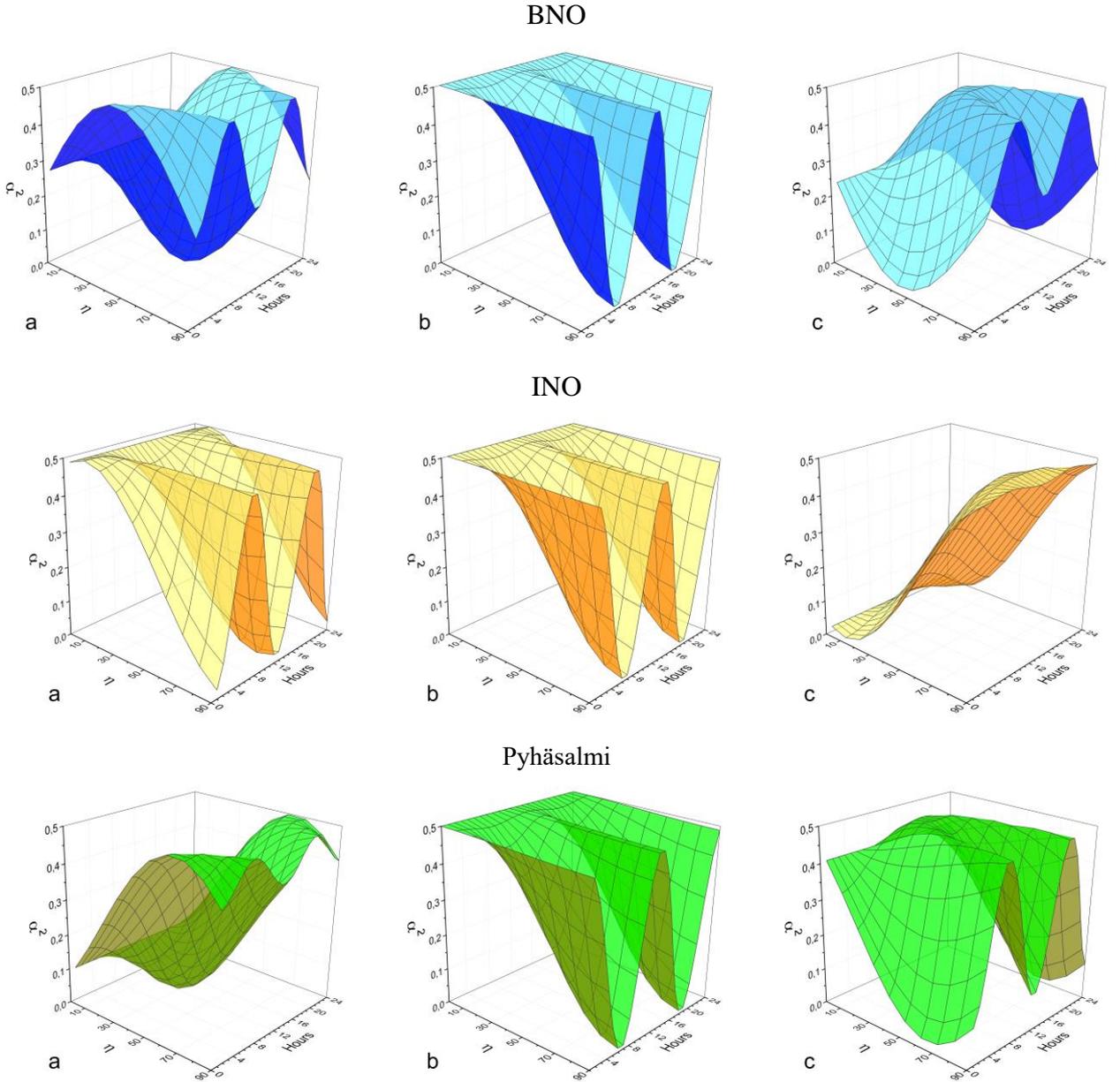

Fig.5. The averaged (for 1 hour) value $\alpha^2 = \langle \cos^2\theta \rangle$ as a function of time and angle η (degrees) for three orientations of the counter: (a) – the vertical one, (b) – along the parallel, (c) – along the meridian, each one - for three mines.

Figures 6 – 8 show strong dependence of the diurnal run of the value $\alpha^2$ from the angle between the direction of the field of dark photons and the axes of the Earth. For each latitude the corresponding curves are given for the angle η = 63º, that corresponds at certain phase to the direction of the field to galactic pole, and the curves for the angle η = 27º, that corresponds at certain phase to the direction in the plane of the galactic disk. One can see substantial difference

of the diurnal run in both cases for three detectors at each latitude that will enable for big enough statistics to determine the absolute phase of the diurnal run and thus to find the absolute direction of the field.

It is very important that we don't need here to measure with high precision the absolute value of the effect to get the interesting result but only to measure how the effect is varied during day. It's all the more important that in experiments searching for dark photons usually it is very difficult to prove that primary ionization gets not partially lost in the interval between when it is originated and when it is detected. In this case we don't have this problem. As one can see also from these figures the diurnal run of the count rate of the counter placed along the parallel for a given direction of the field does not depend upon the geographical latitude. This can be used as a controlling measurement in each mine.

It seems appropriate also to use one more counter with a rough (matte) surface of the cathode to do another controlling measurement. This is necessary to do because the counter can "see" direction only if the surface of the cathode is smooth (mirror-like). The counter with a matte cathode will lose the directionality and we should get another diurnal run of the count rate. As a material of the cathode it looks appropriate to use copper with low concentration of radioactive impurities with the electrodeposited layer of aluminum, chrome or nickel as a reflecting material. To produce a matte surface one can lay the same reflecting material by sputtering. The work function of aluminum is 4.2 eV, the one of chrome - 4.6 eV and one of nickel - 4.91÷5.01 eV. Consequently, these metals have different threshold to emit electron and this also can be used in experiment. The capsule of the detector can be fabricated from titanium. This metal is good from technological considerations and contains relatively low concentration of radioactive impurities as our previous measurements have shown.

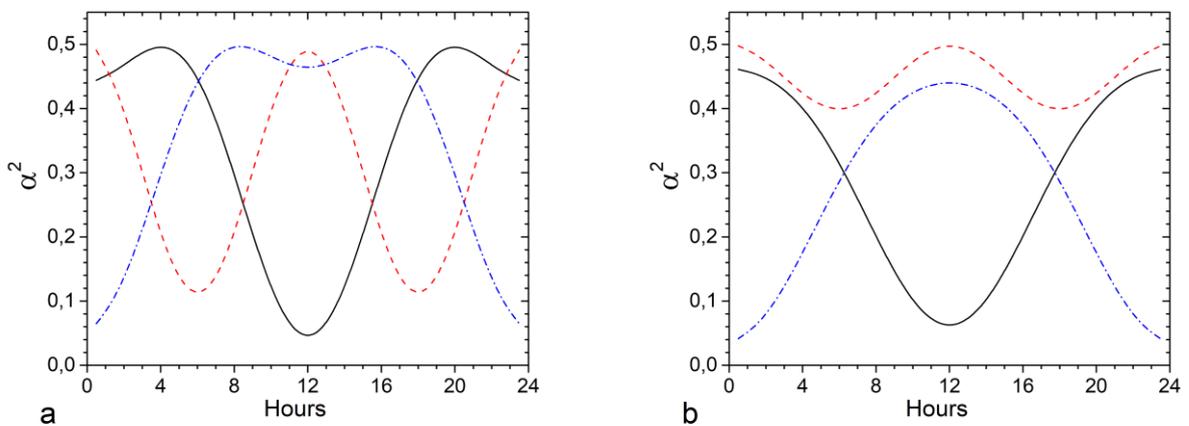

Fig. 6. The averaged (for 1 hour) value $\alpha^2 = \langle\cos^2\theta\rangle$ for BNO, Russia as a function of time for three orientations of the counter: solid line – the vertical one, dashed line – along the parallel, dot-dashed line along the meridian. Left picture: for $\eta = 63°$, right picture: for $\eta = 27°$.

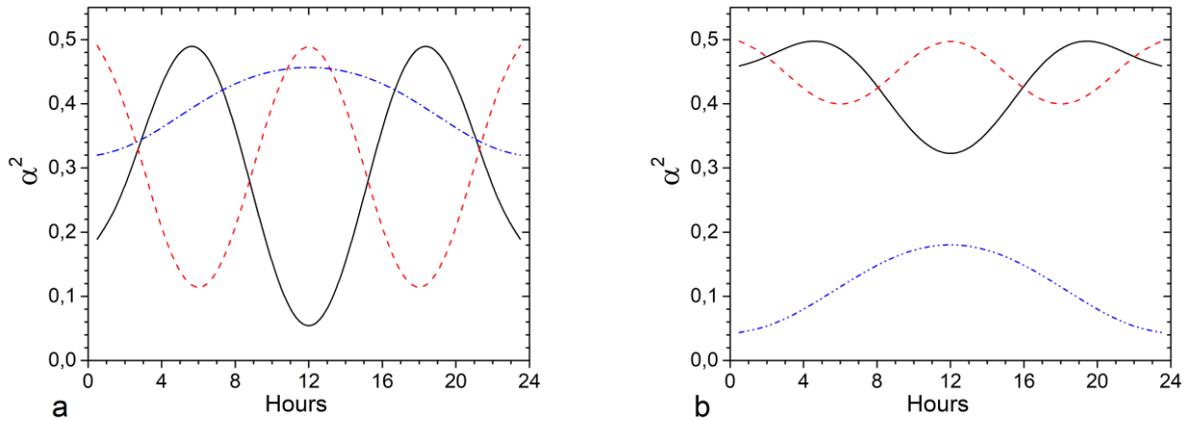

Fig. 7. The same as Fig. 6 but for INO, India

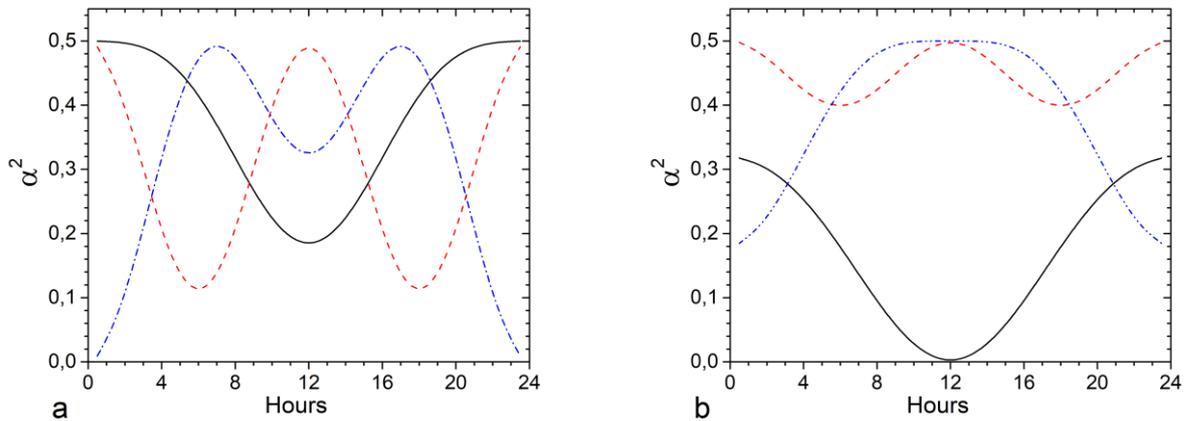

Fig. 8. The same as Fig. 6 but for Pyhäsalmi, Finland.

## CONCLUSIONS

The possibility to conduct the experiment on the search of diurnal variations of the effect from dark photons with three multicathode counters with orthogonal orientation at three mines at different geographical latitude: Pyhäsalmi in Finland, Baksan in Russian Federation and INO in India has been considered. The measurements conducted at three different latitudes in case if the difference of the diurnal run of count rates is observed will enable to get the evidence of the anisotropy in the direction of the field of dark photons by three independent experiments. The important thing is that here is not necessary to measure precisely the effect but only its diurnal variations. It is shown also that if the field of dark photons has a certain direction, its vector can be restored from diurnal runs and the absolute phase of measurements, found from data of this experiment. The important details in the construction and work of these detectors are discussed.

## ACKNOWLEDGEMENTS

The work was supported by the basic research program of INR RAS. We acknowledge useful discussion of this work with G. I. Rubtsov.